# Effects of a Co-Regulation Model for MR Teacher Training: HRV and Self-Compassion as Indicators of Emotion Regulation

Lara Chehayeb, Saarland University, Saarbrücken, s8lacheh@uni-saarland.de
Katarzyna Olszynska, Osnabrück University, Osnabrück, kolszynska@uni-osnabrueck.de
Chirag Bhuvaneshwara, K8 design, Germany, chb@k8.design
Dimitra Tsovaltzi, DFKI GmbH, Kaiserslautern, dimitra.tsovaltzi@dfki.de

**Abstract:** Teachers play a pivotal role in fostering students' emotional and cognitive development. Teachers need to regulate their emotions in order to co-regulate students. Here using a unique mixed method approach, we investigate the relationship between self-compassion, treating oneself with compassion, and physiological stress responses among pre-service teachers. Heart rate variability (HRV) was measured during a mixed reality (MR) teacher training scenario environment designed to simulate socioemotional conflict in class. Recorded interviews that followed the MR-training were analyzed for observed self-compassion. Findings suggest that less emotional stress during the MR-training correlates with higher levels of self-compassion during the interview. MR-trainings and self-compassion may be valuable tools to train teacher emotion regulation and well-being.

## Introduction

The teaching profession is characterized by dealing with complex socio-emotional dynamics. While the primary responsibility is perceived to be the cognitive development of the student, teachers are also highly responsible for the student's social and emotional development. Teachers assume a co-regulating role similar to that of caregivers in the socialization process to effectively manage their classrooms (James, 2011b; Goetz et al., 2021). Ideally, by effectively regulating their own emotions, teachers can create a safe learning atmosphere, allowing them to concentrate on the students and the teaching process. The ability to regulate emotions constructively not only enhances individual emotional experiences of teachers but also promotes effective interpersonal emotion regulation and impacts how situations are handled (Frenzel et al., 2021; Gross, 2015). Sending conflicting signals contradicts the contingency principle for developing emotional regulation (Gergely & Watson, 1996), leading to confusion for students regarding their own emotions and impeding their ability to regulate emotions effectively. Teachers run this risk when they inadvertently send mixed emotional signals as they are unable to automatically regulate their emotional expression, such as feeling anger but trying to stay calm. Consequently, it is crucial for teachers to develop their own emotion regulation skills, which are vital for understanding others' perspectives (Hargreaves, 2000). Although teachers may use explicit strategies to support the socio-emotional well-being of their students, their own implicit emotional awareness significantly influences their ability to conceal or mask their emotions in the classroom (Hargreaves, 2000). To investigate these implicit processes more accurately, we conduct a study within a project, utilizing Mixed Reality technology and physiological measures (Chehayeb et al., 2024).

## Mixed Reality

Mixed reality (MR) environments blend the real and virtual worlds, creating immersive settings for real-time interaction between physical and digital objects, making them particularly advantageous for educational purposes (Thiede et al., 2022). They allow for realistic, hands-on learning experiences that help teachers simulate and practice managing complex socio-emotional challenges in a safe environment. A crucial element of MR environments is socially interactive agents (SIAs), which include embodied conversational agents, intelligent virtual agents, and social robots (Holz et al., 2011). Autonomous agents engage in social interactions and imitate human-like behaviors, providing a dynamic space to train and develop social skills like empathy, emotional control, and conflict resolution (Cassell, 2000). Realistic simulations offer a valuable alternative to the challenges of collecting real-world data, which can be time-consuming, costly, and logistically complex. By utilizing controlled, realistic simulations, it becomes possible to record physiological measures like heart rate variability (HRV), providing rich data to deepen our understanding of socio-emotional experiences.

## Heart Rate Variability

HRV is a widely utilized metric for assessing an individual's responses to stress and emotional challenges. In the context of stress, HRV is typically observed to decrease, which is indicative of an elevated sympathetic (so called 'flight or fight') response and a reduction in parasympathetic ('rest and digest') activity. Conversely, recovery and emotional regulation processes are often characterized by an increase in HRV, which suggests a return to a balanced autonomic state (Steffen et al., 2020). This physiological reactivity, which is also known as "autonomic flexibility" or "vagal flexibility", reflects an individual's ability to engage with, and recover from, stressors (Berntson et al., 2008; Thayer & Lane, 2000). A higher HRV is indicative of greater autonomic flexibility and superior overall health, whereas a lower HRV is associated with stress, anxiety, and emotional dysregulation (Segerstrom & Nes, 2007).

## Self-compassion as an Emotion Regulation

In social contexts, e.g. in a classroom, automatic regulation strategies are essential for facilitating effective interactions, particularly through co-regulation mechanisms. Successful emotion co-regulation requires a state of moderate affective empathy—the effortless ability to understand and validate the emotions of others (Davis et al., 1994). These mechanisms become essential in high-arousal situations, such as conflicts, where emotional awareness is diminished, and explicit regulation strategies may prove challenging or even counterproductive (Järvelä et al., 2019). Reflecting on situations is a critical process for fostering personal and interpersonal growth, particularly in the aftermath of conflict situations. Engaging in self-reflection enhances self-awareness as people tend to understand the reasons behind events or actions by directing attention inward or outward (Fenigstein et al., 1975). Increased self-awareness, in turn, facilitates empathy and perspective taking (Teasdale et al., 2002), and may even counter conflicting emotional signals. In this study, we examine the process of compassionate self-reflection and its intrinsic connection to how individuals experience challenging situations. Specifically, we prompt participants in a reflective interview to revisit the situation and consider their experiences while drawing on their own personal resources.

Self-compassion, an emotion regulation strategy grounded in compassion, provides a valuable approach in such contexts. It involves offering nonjudgmental acceptance and understanding of one's emotions (Neff, 2003), and encouraging individuals to attend to and confront negative emotions with care. By fostering acceptance and understanding of subjective emotional experiences, self-compassion creates a foundation for compassionate behavior toward others (Neff, 2011). Elevated levels of self-compassion enhance psychological well-being, reduce burnout, and decrease levels of depression, anxiety, and perceived stress (Ferrari, M. et al., 2019). Previous research has demonstrated that self-compassion acts as a positive and protective factor, whereas self-criticism increases vulnerability to stress symptoms (Montero-Marin et al., 2016). Furthermore, Lopez et al. (2015) found that the positive components of the Self-Compassion Scale (SCS) are significantly associated with positive affects, further highlighting the beneficial role of self-compassion in emotional well-being. Individuals who adopt a self-compassionate approach when reflecting on experiences are better equipped to navigate challenging emotional contexts, as this approach fosters emotional resilience and promotes adaptive emotion regulation (Gross & John, 2003). The concept of compassion and self-compassion extends beyond just an emotional response, encompassing a range of cognitive and motivational dimensions. The physiological benefits of compassion-based practices, such as Loving-Kindness Meditation (LKM), have been demonstrated by research which has identified the activations of neural circuits associated with caregiving and emotion regulation (Hofmann et al., 2011; Lutz et al., 2008). Longe et al. (2010) discovered that self-compassion engages the brain regions associated with self-soothing and stress reduction, including the insula and anterior cingulate cortex.

In this study, we test the possibility of training teachers' emotion regulation through MR-trainings. We employ a mixed-methods approach, combining qualitative analysis of reflective interviews and physiological measures. Previous research has demonstrated a positive association between self-compassion and mental health, including its ability to mitigate stress and emotion regulation difficulties (Finlay-Jones et al., 2015). Additionally, self-compassion has been linked to physiological markers of emotional flexibility, such as higher heart rate variability (Svendsen et al., 2016). Accordingly, we hypothesize that higher levels of state self-compassion among student-teachers will be associated with reduced physiological stress responses during a simulated teaching scenario, as indicated by increased HRV (hypothesis 1). Furthermore, we expect that participants with higher positive self-compassion scores will demonstrate increased perspective-taking behavior (hypothesis 2a) and decreased non-self-reflective behavior, suggesting that self-compassion may contribute to more adaptive social behaviors (hypothesis 2b).

## Methods

### Participants

The study involved *N*=60 participants who were recruited through seminars at a university. These participants were specifically selected based on their enrollment in higher semesters (3rd semester and above) of teacher training programs, as these individuals represent the target demographic group of the project. Recruiting was carried out via email and in-person announcement during seminars. Participants received two study participation points for their involvement in the study. All participants demonstrated proficiency in German language, as the study was conducted in German for both stages.

### Study Design

Before conducting the experiment, we obtained the necessary ethical approval from the relevant institutional review board. The study was structured in two stages. The first stage involved a classroom simulation where participants prepared and delivered a lesson as substitute teachers to virtual student avatars. The simulation was specifically designed to provide a challenging interaction. During this phase, the participant is asked to place a polar band strap around their chest. The polar band measures HRV in real-time, providing physiological indicators of stress. The second part consisted of a post-interaction interview, during which participants are guided by an interviewer to watch critical moments of their experience in the simulation, focusing on their reflections and emotional responses. The interviewer introduces the session as a safe, non-judgmental environment, encouraging participants to openly reflect on their experiences. The questions were constructed to encourage participants to describe their experience and the motives behind their behavior.

The analysis of the interviews was grounded in a coding framework derived from Self Compassion Scale (SCS) (Neff, 2003). This framework facilitates a structured examination of participants' self-compassion responses during reflective discussions. The SCS scale is used with its relevant constructs as coding categories: self-kindness, self-judgement, common humanity, isolation, mindfulness, and over-identification. These categories served as indicators of how participants reflect on their experiences during the interaction (see Table 1), allowing for structured categorization of responses, ensuring consistency in coding across the dataset.

**Table 1**
*Coding categories, definitions and anchor examples for qualitative coding of self-compassion*

| Categories | Definitions | Anchor examples |
|---|---|---|
| **Positive subcomponents** | | |
| Self-kindness | acceptance of emotions and experiences | ""It's okay, I am still learning and improving." |
| Common humanity | acknowledged a shared experience with others, such as recognizing that the challenges faced are common in similar situations | "I know other teachers face similar situations." |
| Mindfulness | balanced awareness of one's emotions, neither suppressing nor becoming overwhelmed | "I noticed I was getting upset but I tried to stay calm." |
| **Negative subcomponents** | | |
| Self-judgement | harsh self-criticism or self-blame | "I feel that I failed." |
| Isolation | expressed feeling of being alone or uniquely burdened by the situation | "I must be the only one that couldn't complete the task." |
| Over-identification | excessive identification with emotions, becoming overly consumed by feelings | "I can't stop thinking about how I behaved in this task." |
| Perspective-taking | understanding and consideration of another person's feelings, thoughts and viewpoint | "I can see why he may be feeling stressed, maybe he's dealing problems home" |
| Non-self-reflective | avoiding considering their own actions or emotions, focusing on external factors | "I could not finish the lesson because of his behavior" |

The coding process was conducted using the ELAN software, which facilitated the annotation of each participant's interview (see Figure 1), with 52 out of the 60 available videos coded. Eight videos were excluded due to inferior quality or missing Polar Band data for respective participants. The ELAN tool allowed for precise annotation (Wittenburg et al., 2006) of specific occurrences in the video, enabling a detailed analysis of the participants' self-compassion behaviors in response to their MR teaching experience.

**Figure 1**
*Example of ELAN Annotation Video*

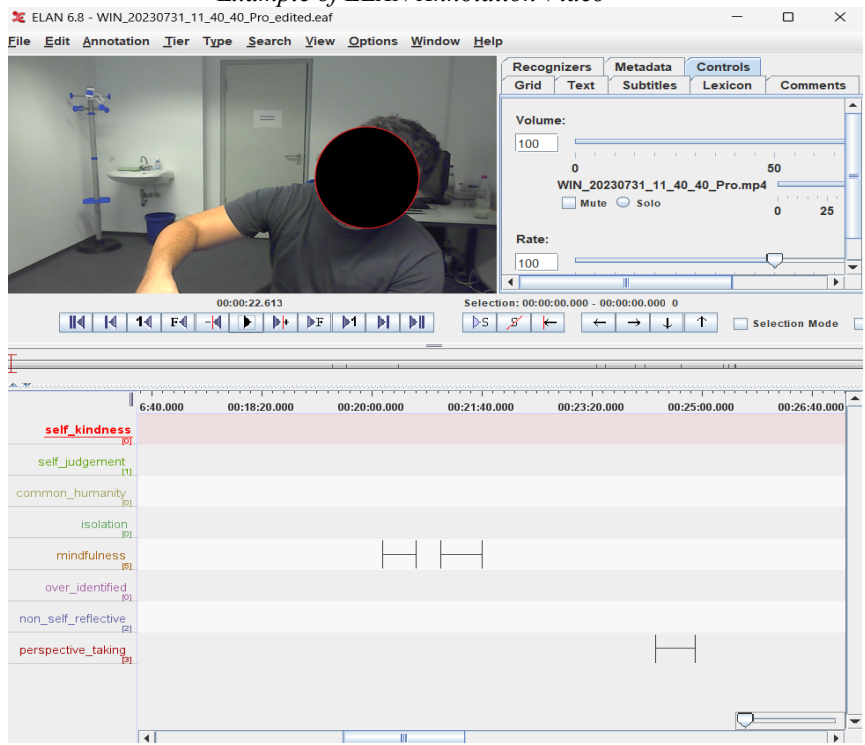

The framework method (Gale et al., 2013) was applied as a rigorous analytical tool that systematically organizes and manages qualitative data. This method supports a deductive approach based on a predefined coding framework, such as Neff's self-compassion scale, which served as a reliable and well-established framework for the analysis. The framework method allows for re-coding and refinement as the data are reviewed, ensuring that the coding process remains consistent and accurate throughout. Through this iterative process, coding was reviewed and adjusted to ensure consistency across all videos, thereby increasing the robustness of the analysis. In addition to the framework method, the Apprenticeship Model (Renkl, 2001) was utilized, emphasizing the importance of learning through close collaboration and shared practice. Approximately 15% of the videos were coded together, with another researcher reviewing and discussing any discrepancies until agreement was reached on the applied codes. This model emphasizes in-depth training, discussion, and agreement between a novice and an experienced researcher, reinforcing the reliability and validity of the coding process. To further ensure the reliability and objectivity of the coding process, a reflexive approach was adopted, with detailed notes taken through the coding. Reflexive documentation is a key strategy for maintaining objectivity and transparency in qualitative research, allowing the researcher to reflect on the decisions made during the coding process (Finlay, 2002). The combined use of the framework method and the apprenticeship model created a robust and reliable process that ensured the accuracy and consistency of the coding.

## Data Processing

### HRV Data Processing

In this study, the relationship between HRV metrics and self-compassion behavior was explored to assess participants physiological stress responses during MR sessions. The HRV metrics: Root Mean Square of Successive Differences (RMSSD), STandard Deviation of Normal-to-Normal Interbeat Intervals (SDNN), and the LF/HF ratio were chosen for their ability to capture both parasympathetic and sympathetic nervous system activity, reflecting emotional regulation, stress, and overall autonomic balance (Laborde et al., 2017; Malik et al., 1996). RMSSD and SDNN, in particular, offer sensitive measures of autonomic regulation, making them preferable over simpler heart rate metrics, as they provide a more accurate representation of physiological stress. The LF/HF ratio, a critical frequency-domain metric, was included due to its established link with stress, where higher ratios are associated with increased sympathetic activation (Shaffer & Ginsberg, 2017). To quantify overall stress, HRV averages for each participant were computed over the interaction phase, which lasted approximately four to five minutes. This approach enabled the study to capture physiological responses during moments of stress without requiring baseline data, which were unavailable for comparison. Missing or incomplete HRV data were addressed by excluding invalid points, ensuring that the frequency-domain metrics could still be calculated meaningfully.

The HRV data processing began with parsing the XML files generated by the Polar Band, which stored electrocardiogram (ECG) data for each participant. Following this, R-peaks representing individual heartbeats were detected in the ECG signal. Consequently, time domain metrics (RMSSD, SDNN) and frequency-domain metrics (LF/HF ratio) were computed for each participant and stored in a CSV file.

## ELAN Data Processing

Self-compassion data was generated from coded video interviews. Each video was coded based on moments where participants reflected on their MR session experiences, specifically related to stressful conflict situations. These coded occurrences were exported to a CSV file and categories such as self-kindness or self-judgement were aggregated into positive and negative self-compassion instances without including non-self-reflection and perspective taking in this overall score. By categorizing self-compassion into these two distinct groups, we aimed to capture the dichotomy between self-compassion and self-criticism, which is highly relevant to stress-related outcomes.

## Data Analysis

To investigate the relationship between HRV metrics and positive self-compassion, correlation analysis was conducted, focusing on RMSSD, SDNN, and the LF/HF ratio- established indicators of autonomic nervous system activity and stress regulation. The analysis sought to determine whether detected positive self-compassion instances were associated with improved physiological stress regulation as reflected by these metrics. As we prompt participants to revisit the situation by watching themselves in the situation and consider their experiences, we gain a better understanding of their subjective experience.

In addition to the physiological measure, the relationship between self-compassion and coded behavioral annotations, including perspective-taking and non-self-reflective behaviors, was analyzed. Correlation analysis was conducted to investigate whether individuals with higher levels of self-compassion engaged in more perspective-taking and fewer non-reflective behaviors.

Prior to conducting correlation analysis, the distribution of the data was assessed using the Shapiro-Wilk test to determine normality. This evaluation was crucial for selecting appropriate correlation methods based on distributional assumption. The Shapiro-Wilk test was chosen due to its superior power in detecting deviations from normality compared to other common tests (Korkmaz & DemiR, 2023).

## Results

52 participants, 14 male and 38 female, with a mean age of $M = 23.9$ years, $SD = 4.97$ were considered for analysis. The results of the Shapiro-Wilk tests suggest that Average_RMSSD and Average SDNN are normally distributed, but both LF/HF ratio and positive compassion score deviate from normality. Consequently, non-parametric methods such as Spearman's correlation were used to analyze the data. This combination was most appropriate for our analysis, as it allowed us to confirm and validate the findings across different statistical assumptions, ensuring a rigorous evaluation of the association (Rovetta, 2020).

As shown in Table 2, the correlation analysis between HRV metrics and positive self-compassion (hypothesis 1) shows that:

**Table 2**

*Correlation between HRV Metrics and Positive Self-Compassion*

| HRV metric | *p*-value | Spearman Correlation *r* |
|---|---|---|
| RMSSD | .332 | .19 |
| SDNN | .096 | .32 |
| LH/HF Ratio | .035 | -.27 |

- RMSSD and Positive Self-compassion: Spearman's correlation for RMSSD was weak ($r = .19$), suggesting that there is no strong relationship between RMSSD and positive self-compassion in this sample
- SDNN and Positive Self-compassion: the correlation was slightly stronger ($r = .32$), but this result does not definitely confirm the hypothesis of a significant relationship between these variables
- LF/HF Ratio and Positive Self-compassion: the Spearman correlation showed a moderate negative correlation ($r = -.27$). This result supports the hypothesis that higher stress levels (as indicated by higher LF/HF ratio) are associated with lower positive self-compassion scores.

These results suggest that while RMSSD and SDNN may not show strong associations with positive self-compassion detected instances, the LF/HF ratio provides meaningful insights into the stress-compassion relationship, particularly highlighting that individuals with higher stress levels exhibit lower levels of positive self-compassion.

The correlation analysis between positive self-compassion and two behavioral indicators: perspective-taking and non-self-reflective (hypothesis 2), show (see Table 3):

**Table 3**

*Correlation Between Positive Self-Compassion and Behavioral Indicators*

|  | *p*-value | Spearman Correlation *r* |
| --- | --- | --- |
| Perspective-taking | <.001 | .38 |

- Positive Self-Compassion and Perspective-taking (hypothesis 2a): Spearman's correlation, testing the monotonic relationship between positive self-compassion and perspective taking showed a moderate positive relationship ($r = .38$), supporting the idea that self-compassion fosters behaviors that align with perspective taking. This indicates that teacher trainees with higher self-compassion scores were more likely to take the perspective of the students, confirming our hypothesis and similar to previous research in emotion regulation (Boland et al., 2021).

Positive Self-compassion and non-self-reflective Behavior (hypothesis 2b): as the instances for non-reflective-behavior were very few, the analysis was not possible. This may have been due to the focused questions on the motives and feelings of the participants, emphasizing a more analytic way of thinking, rather than a more mindful thinking which might have fostered self-compassion. This was a risk we had to take in designing the study in view of understanding motives for future automatic modelling of the teacher co-regulation behavior (Bhuvaneshwara et al., 2023).

## Discussion

This study investigates the relationship between heart rate variability (HRV), a physiological marker of stress, and self-compassionate behaviors, including self-kindness, common humanity, and mindfulness, during a reflective interview. To assess the influence of self-compassion on both physiological and psychological responses, we employ a mixed-methods approach. By combining qualitative analysis of interview data with quantitative analysis of HRV metrics, we aim to gain a deeper understanding of the complex relationship between self-compassion and stress regulation. Preliminary findings suggest that state self-compassion may influence physiological stress responses. The statistically significant inverse relationship between the LF/HF ratio and self-compassion ($p = .035$) supports the idea that self-compassionate individuals experience lower sympathetic arousal (i.e., less stress). This has also been supported by previous research where compassion can positively impact physiological stress responses (Di Bello et al., 2021; Duarte, J. et al., 2016).

The moderate positive correlation between self-compassion and perspective-taking ($p < .001$) aligns with previous research (Gilbert, 2014; Park & Tsovaltzi, 2022). Individuals demonstrating self-compassion are more likely to engage in behaviors that facilitate emotion regulation, such as perspective-taking. This is particularly beneficial in challenging classroom situations, as teachers can employ conflict resolution strategies that empathetically engage students while achieving pedagogical goals. Such approaches may contribute to a positive learning environment and mitigate teacher burnout.

The correlation analysis revealed no significant associations between RMSSD, SDNN, and self-compassion. This lack of significant findings may be attributed to the fact that the induced stress was primarily concentrated at the beginning of the simulation. As no further stressors were introduced after the initial peak, participants may have had more time for physiological recovery through 'rest and digest'. This may have limited the variability in HRV metrics, making it difficult to detect subtle differences associated with self-compassion. Future research is needed to investigate the relationship between experienced stress and RMSSD and SDNN under a wider range of conditions.

The scarcity of non-reflective behaviors observed in the post-interaction interviews likely reflects the nature of the interview structure, which emphasized introspection on participants motives and feelings. This design may have inadvertently constrained opportunities for participants to adopt a broader perspective by considering the virtual student's behavior or situation, as well as a self-compassionate, more mindful approach towards themselves.

From a physiological standpoint, HRV metrics, including RMSSD, SDNN, and LF/HF ratio, were employed to evaluate participants' stress responses during MR sessions. These metrics, averaged over the interaction phase, provided a reliable overview of autonomic nervous system activity in the absence of baseline measures. While this approach enables meaningful comparisons, it underscores the importance of integrating longer, more varied interventions and capturing baseline HRV data to better contextualize stress responses. The exclusion of invalid data points and the reliance on frequency-domain metrics like LF/HF ratio allowed the study to account for variability across participants while addressing challenges posted by available data. Furthermore, a larger sample could provide a more robust analysis of the relationship between self-compassion and HRV metrics. Additionally, difficult as it may be, providing a baseline for HRV prior to the MR sessions, would have allowed determining whether observed HRV changes were due to the interventions itself or pre-existing individual differences. Incorporating baseline HRV measurements in future studies would provide a clearer picture of the participants physiological responses to stress.

## Conclusion

Addressing teacher stress is crucial for ensuring the sustainability of the teaching profession and preventing adverse health consequences. MR simulations offer a valuable tool for replicating authentic classroom scenarios, enabling the exploration of pedagogical strategies and the development of essential professional skills. The findings of this study highlight the significance of self-compassion as a critical socio-emotional skill for managing stress during challenging teaching situations. Self-compassion may play an important role not only in increasing the overall job satisfaction and performance but also enhancing teacher well-being.

While this study provides valuable insights into the potential relationship between self-compassion and stress regulation, future research could benefit from real-time HRV data collection during reflective interviews. This would allow for a better understanding of the immediate physiological impact of self-compassion.

## Acknowledgments

This study is funded by the German Federal Ministry of Education and Research within the funding line "Interactive systems in virtual and real spaces - Innovative technologies for the digital society" (Project MITHOS, grant 16SV8687).